# Opposing Shear-Induced Forces Dominate Inertial Focusing in Curved Channels and High Reynolds Numbers

Eliezer Keinan[1], Elishai Ezra[1], Yaakov Nahmias [1,2, *]

1. Alexander Grass Center for Bioengineering, Benin School of Computer Science and Engineering, The Hebrew University of Jerusalem, Jerusalem, ISRAEL
2. Department of Cell and Developmental Biology, Silberman Institute of Life Sciences, The Hebrew University of Jerusalem, Jerusalem, ISRAEL

* Correspondence should be address to:
  Prof. Yaakov Nahmias
  E-mail: ynahmias@cs.huji.ac.il

## Abstract

Inertial focusing is the migration of particles in fluid toward equilibrium, where current theory predicts that shear-induced and wall-induced lift forces are balanced. First reported in 1961, this Segre-Silberberg effect is particularly useful for microfluidic isolation of cells and particles. Interestingly, recent work demonstrated particle focusing at high Reynolds numbers that cannot be explained by current theory. In this work, we show that non-monotonous velocity profiles, such as those developed in curved channels, create peripheral velocity maxima around which opposing shear-induced forces dominate over wall effects. Similarly, entry effects amplified in high Reynolds flow produce an equivalent trapping mechanism in short, straight channels. This new focusing mechanism in the developing flow regime enables a 10-fold miniaturization of inertial focusing devices, while our model corrects long-standing misconceptions about the nature of mechanical forces governing inertial focusing in curved channels.

## Introduction

Inertial focusing is the migration of particles across streamlines due to differences in lift force acting on the particle surface. The phenomena has numerous applications in micro-particle manipulation ranging from microfluidic cell sorting to particle mixing and ordering[1, 2]. First reported by Segre and Silberberg in 1961[3], a full analytical solution of the forces that dominate particles in Poiseuille flow was provided by Ho and Leal thirteen years later[4]. Ho and Leal showed that particles migrate from the center of a channel towards the wall due to shear-induced lift forces, and are rejected from the channel perimeter by wall-induced lift forces creating a stable equilibrium at a distance of $0.6 \times R$ from the center of the channel[4].

Applications of inertial focusing soared in the advent of microfluidics especially in curved channels where Dean forces could be used to enhance sorting accuracy[1, 2, 5]. Current understanding of sorting in curved channels rely on Ho and Leal description of a balance between shear-induced and wall-induced lift forces, suggesting that focusing is limited to low Reynolds (Re) numbers [6, 7]. Surprisingly, recent work demonstrated inertial focusing at high Re numbers, where wall-induced lift forces are negligible[8-10], demonstrating a failure of current understanding. Therefore, there is a need to elucidate the mechanism that dominates inertial focusing at high Re numbers[1, 2].

Recently, Ciftlik and colleagues exploited multilayer fabrication to explore inertial focusing at high Re numbers[6]. Surprisingly, particles focused rapidly at high flow rates, requiring only a short channel footprint. However, while increasing fluid velocity at low Re numbers (Re < 150) pushed particles toward the channel walls, as suggested by current theory[8], at high velocities (Re > 300) particles migrated toward the center of the channel[6]. Interestingly, a similar signature can be seen in the experiments of Matas and colleagues reported over a decade ago[11]. Quantifying the effect of *tubular pinch* in macroscale tubes, they showed particles moved toward the wall with increasing Re (100 < Re < 700). However, at high velocities (Re =1000), a secondary equilibrium appeared at 0.5×R, with a single inner equilibrium at Re = 1650. Each of these findings is considered to be a failure of the current theory[11, 12].

In this work we demonstrate that at high Re numbers focusing takes place far from the channel walls due to opposing shear-induced lift forces, formed around a peripheral velocity maxima. Such peripheral velocity maxima can form in curved channels, even in low Re numbers, suggesting that the prevalent explanation for inertial focusing in curved channels is in error[13-16]. In addition, our results suggest that entry effects can dominate focusing behavior in short straight channels at high Re number, producing the framework for 10-fold miniaturization of inertial focusing devices.

**Theoretical Background**

Shear-induced lift forces are caused by a velocity gradient impinging across a particle width (**Fig. 1A**). Pressure increases where fluid velocity is greater than particle velocity, pushing particles down the velocity gradient toward lower pressure. Ho and Leal described forces acting on small rigid spheres in low Re numbers using Lorentz generalized reciprocal theorem[4]. Their analysis neglected forces originating from lag velocity or the rotation slip of the particles, as these are orders of magnitude smaller than the stresslet contribution. The resulting general force equation includes wall-induced lift force and shear-induced lift force and is given by:

$$(1) \quad F_L = \kappa^2 Re \, [\beta^2 G_1(s) + \beta \gamma G_2(s)]$$

where $\kappa$ is the ratio of particle to pipe diameter, $\beta$ is the local shear rate, $\gamma$ is the shear rate gradient, and s is the radial location of the particle ranging from 0 to 1. $G_1$ and $G_2$ are the position-dependent integration constants for a first order Bessel function[4, 17]. This general equation can be simplified in Poiseuille flow to show that the Segre-Silberberg is a result of a stable equilibrium at 0.6×R resulting from a balance between shear-induced and wall-induced lift forces (**Fig. 1A,B**)[3].

**Non-Monotonous Flow**

Entry-effect-induced destabilization of fluid velocity profile was studied theoretically and experimentally in a variety of applications[18, 19]. It was shown that friction slows fluid near channel walls causing a separation of flow regimes. Flow separation causes fluid velocity to increase adjacent to the wall due to mass conservation[18], resulting in peripheral velocity maxima and a saddle shaped velocity profile (**Fig. 1D**). This velocity profile eventually stabilizes into Poiseuille flow in the laminar flow regime.

Stabilization distance increases with the Re number. Durst and colleagues[20] used numerical simulations to show that length required for stabilization to 99% correlation with parabolic velocity profile ($L_{spp}$) is:

$$(2) \quad L_{spp} = D \cdot [(0.62)^{1.6} + (0.06 \cdot Re)^{1.6}]^{\frac{2}{3}}$$

Using a similar methodology, we can show that a the length required for 95% stabilization in a rectangular channel is given by:

$$(3) \quad L_{spp} = 0.6 \cdot D(Re/10 - 1)$$

We note that velocity profile in curve channels resembles the saddle-like shape of disturbed flow due centrifugal forces pushing fluid toward the concave side of the channel[21] (**Fig. 1C**).

# Results

*Inertial Focusing in Curved Channels*
Pioneering work of Di Carlo and others experimentally demonstrated rapid focusing of large particles in curved channels due inertial effects[22][23]. While the Segre-Silberberg is routinely evoked as explanation, fluid velocity profile is far from parabolic (**Fig. 1C**) suggesting an alternative explanation is in order.

To address this question, we fabricated high aspect ratio channels that are 33-μm high, 200-μm wide, and have a radius of curvature of 900 μm (**methods**). Particles in curved channels experience Dean drag force in addition to the inertial lift, and the ratio between forces ($R_f$) is given by[24]:

$$(4) \quad R_f = \frac{2r \cdot a^2}{D_h^3}$$

where r is radius of curvature, a is the particle diameter and $D_h$ is the channel hydraulic diameter. To minimize Dean drag force-derived migration in our analysis, we used 15.5-μm diameter particles resulting in $R_f$ value of 2.33 a regime in which inertial lift forces dominate.

We chose to focus our analysis in the intermediate flow regime where two distinct focusing behaviors were observed. At Re numbers 43 and 86, a single dominant focus point was observed close to the concave edge of the channel (**Fig. 2A,B**, *orange arrows*). However, at Re numbers 229 and 257, two non-equal focusing points were observed with the dominant point close to the convex edge of the channel (**Fig. 2C,D**). To understand this behavior, we used finite element modeling to characterize the fluid velocity profile in each experimental condition. Our analysis shows a global fluid velocity maximum dominating the concave side of curved channels in all conditions. However, at Re number greater than 150, a secondary fluid velocity maximum appeared on the convex side of the channel (**Fig. 2**).

Using Ho and Leal's general solution [Eq. 1] we derived the force profile based on the numerically derived velocity profile. Our model critically identified the focusing points in all four experimental conditions (**Fig. 2A-D**, *orange arrows*). We show a new equilibrium point on the inner edge of the global velocity maximum formed by opposing shear-induced lift forces precisely matching the focus at Re numbers 43 and 86 (**Fig. 2A-B**, *orange arrows*). The emergence of a second velocity maximum at Re numbers 229 and 257 results in two additional equilibrium points at the convex side of the curve precisely matching experimental results (**Fig. 2C-D**, *orange arrows*). Finally, a weak equilibrium appears close to the middle of the channel at Re number of 257 and is seen as a weak streak in the experimental image (**Fig. 2D**). Taken together, the results clearly show that a simple balance of lift forces applied on a numerically derived velocity field can predict inertial focusing in curved channels.

Remarkably, in contrast to current understanding, wall-induced lift forces don't play a role in the formation of these traps even at low Re numbers. Removing wall-induced lift forces from Ho and Leal's solution produces no difference in trap location (*data not shown*).

### *Critical Values for Internal Focusing in Non-Monotonous Flow*
Previously, Baghat and colleagues calculated that the channel length required for inertial focusing ($L_{fcs}$) is given by[25]:

$$(5) \quad L_{fcs} = D_h \cdot \frac{3\pi}{2} \cdot \frac{1}{Re} \cdot \left(\frac{1}{\kappa}\right)^3$$

where $D_h$ is the channel hydraulic diameter and $\kappa$ is the ratio of particle to channel diameter. The solution assumes that particle migration is dissipated by Stokes drag. We note that stabilization length [Eq. 3] is proportional to the Re number, while the minimal focusing length [Eq. 5] is inversely proportional to it. Equating equations 3 and 5 allows us to calculate a critical Re number for which entry-effects dominate the focusing phenomena:

$$(6) \quad Re_{crit} = 37\kappa^{-1.42}$$

For values of $Re/Re_{crit}$ higher than unity, we predict entry effects will dominate inertial focusing.

### *Inertial Focusing in Short Rectangular Channels*
Recently, Ciftlik and colleagues demonstrated inertial focusing at Re numbers up to 1500 using multilayer metal-oxide fabrication of rectangular channels[6]. Channels were 50-μm wide, and 80-μm high, while fluorescent beads were 10-μm in diameter.

Again, we chose to focus our analysis on the intermediate flow regime where two distinct focusing behaviors were observed. At Re numbers 150 and 450, focusing points were shown on the edge of the channels suggesting classical Segre-Silberberg effect (**Fig. 3A-B**). However, at Re numbers 750 and 1080, focusing points shifted toward the middle of the channel while wall-induced lift forces were expected to diminish (**Fig. 3C-D**). To understand this behavior, we used finite element modeling to characterize the fluid velocity profile in each experimental condition. Our analysis shows a central velocity maximum dominating flow at Re 150 in both axis and in the short axis of Re 450 (**Fig. 3A-B**). However, higher Re numbers increased stabilization distance causing the velocity profile to separate into a characteristic saddle shape (**Fig. 1D**, **3C-D**)

Again, we used Ho and Leal's general solution [Eq. 1] to derive the force profile based on the numerically derived velocity profile. Our model critically identified the focusing points in all four experimental conditions (**Fig. 3A-D**, *arrows*). At Re number 150 the entry effect parameter Re/Re$_{crit}$ is 0.31 and a parabolic velocity profile dominates in both axes. Inertial focusing produces four traps at the edges of the two central axes of the rectangular channel in a diamond pattern (**Fig. 3A**). These are classical Segre-Silberberg equilibrium points at locations where wall-induced forces balance shear-induced forces (*green arrows*). However, at Re number 450 Re/Re$_{crit}$ is close to 1. Under these conditions, a velocity saddle appears along the vertical axis, while the shorter horizontal axis remains parabolic (**Fig. 3B**). As expected, classical equilibrium appears at the edges of the parabolic horizontal axis, where wall-induced balance shear-induced lift forces (*green arrows*). However, entry effects dominate the vertical axis, producing traps closer to the center of the channel (*orange arrows*) on the inner edge of the velocity maximum. These traps are the result of opposing shear-induced lift forces.

Interestingly, while Ciftlik assumed that the diamond trapping pattern was maintained at Re of 750 and above[6], our model demonstrates that entry effects form two velocity saddles, along both horizontal and vertical axes, shifting traps to a rectangular pattern (**Fig. 3C,D**). At Re number of 750 the Re/Re$_{crit}$ is 1.53 which is sufficient to produce opposing shear-induced lift force traps on both axes (*orange arrows*). We note that force equilibriums in the middle of the channel are unstable, while those balanced by wall-induced lift forces are too close to the channel walls to trap a spherical particle (**Fig. 3C**). Re number of 1080, Re/Re$_{crit}$ of 2.21, producing a similar rectangular pattern, although experimental disorder appear to bias particles to one side or the other (**Fig. 3D**).

## Discussion

Until recently, inertial focusing applications were restricted to relatively low fluid velocities limiting the throughput of each microfluidic channel [22, 26-28]. This restriction was a result of technical difficulties in the fabrication of microfluidic devices that can sustain high pressures[29, 30], and an inadequate understanding of the mechanism of focusing at high Re numbers[1, 2, 6].

In this work, we relied on adhesive tape cleaning and a fluoroalkyl trichlorosilane (FTS) vapor environment to form strong covalent bonds between PDMS and glass, sustaining flow up to Re number 300[14, 29]. This fabrication protocol allowed us to demonstrate inertial focusing in curved channels at relatively high fluid velocities (**Fig. 2**). Our results join a rapidly expanding body of literature demonstrating particle focusing in straight channels at high fluid velocities[6, 12, 17]. In fact, inertial focusing was demonstrated at Re number up to 1650 [6, 11]. However, while the basic phenomena of focusing at low Re numbers is ubiquitously understood to be the result of a balance between shear-induced and wall-induced lift forces (**Fig. 1A-B**), there is little understanding of the focusing phenomena in the intermediate flow regime [1, 2, 6].

Our results demonstrate that saddle-like velocity profiles, such as those produced in curved channels or by entry effects (**Fig. 1C-D**) form focusing points on the inner edge of the velocity maximums, close to the channel center. This focusing behavior is governed by opposing shear-induced lift forces and can be derived by applying Ho and Leal's general solution [Eq. 1] on numerically derived velocity profiles. In fact, recent work by Hood and colleagues expands Ho and Leal's theoretical analysis to high Re and three-dimensional flow, validating our approach by demonstrating that the general solution is valid far beyond its original assumptions[31, 32].

Importantly, our insight into inertial focusing in curved channels should correct long-standing misconceptions in the field. It is clear that opposing shear-induced lift forces dominate focusing in curve channels, even at low Re numbers, while the Segre-Silberberg effect does not appear to play a role in this geometry (**Fig. 2**). Focus points are closer to the channel center and shift between the concave to convex edge with increasing Re numbers.

Interestingly, the developing flow regime is often neglected in microfluidics due to the low Re numbers involved. Past theoretical analyses of inertial focusing assumed fully developed flow is responsible for focusing[11] and parabolic boundary conditions were applied in simulations to avoid entry effects[33]. In fact, recent work simulating velocity profiles in focusing devices used a flat entry velocity profile, ignoring the peripheral velocity maxima and thus failed to observe our phenomena[34]. Our results suggest that opposing shear-induced lift forces should dominate for $Re/Re_{crit} > 1$ where the focusing path is shorter than the fluid stabilization distance, while values of $Re/Re_{crit} < 1$ should reproduce the classical Segre-Silberberg effect. Therefore, focusing devices working at high Re numbers can be 10 to 100-fold shorter than current designs, only limited by focusing length [Eq. 5].

In summary our work provides a framework to predict inertial focusing location in microfluidic devices at high Re numbers. The combination of lift forces calculated using our framework with a size-dependent mechanism such as Dean force would lead to more precise particle sorting and a simple a priori design of high throughput inertial focusing devices.

## Materials and Methods

*Materials*

PDMS was purchased from Dow Corning (Midland, MI). SU8-3025 was purchased from Microchem (Newton, MA). 15.5-µm diameter fluorescent yellow-green particles were purchased from Bangs labs (Fishers, IN), and Pluronics F68 was purchased from Sigma Aldrich (Rehovot, Israel).

*Numerical Simulations*

Numerical simulations were performed using the COMSOL MultiPhysics simulation platform v4.3b with a direct linear system solver and extra fine mesh. Fluid density was defined as $1\times10^3$ kg/m$^3$ with a dynamic viscosity of $1\times10^{-3}$ Pa·s. Cross sections are shown from the middle of the curved channel (**Fig. 2**) and $4\times D_h$ from the rectangular channel entrance (**Fig. 3**). Velocity profiles were extrapolated to 9$^{th}$ grade polynomials using MATLAB to calculate the first ($\beta$) and second ($\gamma$) fluid velocity derivatives [Eq. 1].

*Device Fabrication*

Microfluidic devices were fabricated by soft lithography. Briefly, molds were fabricated by photolithography of SU8 on silicon wafers at the Harvey Krueger Center of Nanoscience and Nanotechnology at the Hebrew University of Jerusalem. Channels were replica molded in PDMS and bonded to glass using oxygen plasma as previously described [32, 35, 36]. Channels and glass were cleaned using 3M low-residue adhesion tape (St. Paul, MN). Prior to the experiment, the channel was coated with Pluronic F-68 for 1 hour at room temperature to prevent non-specific adhesion of fluorescent particles. Microbeads were perfused using a Fusion 200 syringe pump (Chemyx, Stafford, Texas) and imaged on a Zeiss Axiovert Microscope.

## Acknowledgements

This work was funded by ERC Starting Grant TMIHCV (N° 242699) and the Marie Curie Reintegration Grant microLiverMaturation (N° 248417).

# Figure Labels

**Fig. 1. Modeling entry effect derived inertial focusing.** (A) Schematic depicting shear-induced lift force in classical Poiseuille flow. (B) Parabolic velocity profile at Re number 30. Stable equilibrium points appear at the edges of the profile due to opposing wall-induced and shear-induced lift forces. (C) Saddle-shaped velocity profile produced by centrifugal forces in curved channels at Re number 229. A new equilibrium point appears on the inner edge of the global velocity maximum closer to the channel middle. A second equilibrium point appears on the inner edge of the local velocity maximum on the convex side of the channel. Traps closer to the wall disappear, as they cannot accommodate large particles. (D) Saddle-shaped velocity profile at Re number 1000 produced in developing flow regime due to entry effects. New equilibrium points appear close to the center of the channel due to opposing shear-induced lift forces.

**Fig. 2. Forces governing inertial focusing in curved channels.** Experimental trapping of 15.5-μm beads in curved channels fabricated using soft lithography (*center*). Numerically calculated fluid velocity profiles matching experimental conditions (*left, bottom*). Analytical force calculation based on modeled velocity profile for horizontal cross-sections (*white dashed lines*). All equilibrium points identified are due to opposing shear-induced lift forces and marked by orange arrows. (A) Re number 43 focusing near concave side of the channel. (B) Re number 86 focusing near concave side of the channel. (C) Re number 229, two additional equilibriums emerge on convex side of the channel due to a local velocity maximum. (D) Re = 257, an additional equilibrium appears at the center of the channel as predicted by the force analysis.

**Fig. 3. Modeling inertial focusing in rectangular channels.** Experimental trapping of fluorescent beads in linear channels reproduced with permission from *Ciftlik et al.* [6] (*left*). Numerically calculated fluid velocity profiles matching experimental conditions (*middle*). Analytical force calculation based on modeled velocity profile for horizontal and vertical cross-sections (*white dashed lines*). Equilibrium points generated by opposing wall and shear-induced lift forces marked by *green arrows*, while those due to opposing shear-induced lift forces marked by *orange arrows*. (A) Re number 150 ($Re/Re_{crit}=0.31$) only near wall equilibrium appear. (B) Re number 450 ($Re/Re_{crit}=0.92$) centrally located equilibrium points appear on the vertical axis due to peripheral velocity maxima. (C) Re number 750 ($Re/Re_{crit}=1.53$), centrally located equilibrium points appear in both axis. (D) Re number 1080 ($Re/Re_{crit}=2.21$), centrally located equilibrium points appear in both axes. Horizontal focus shifts toward wall in both model and experiment.

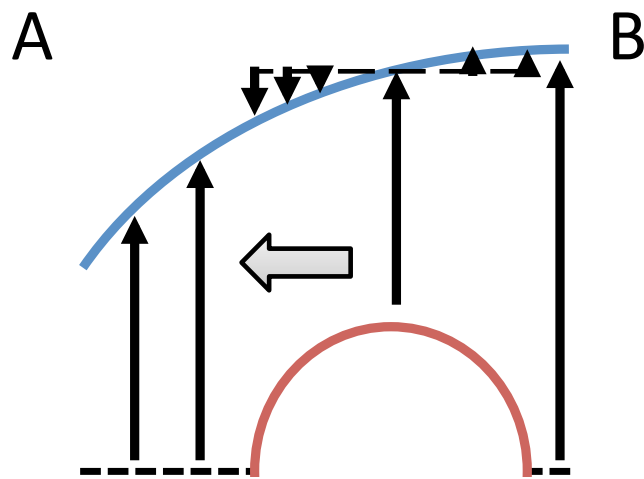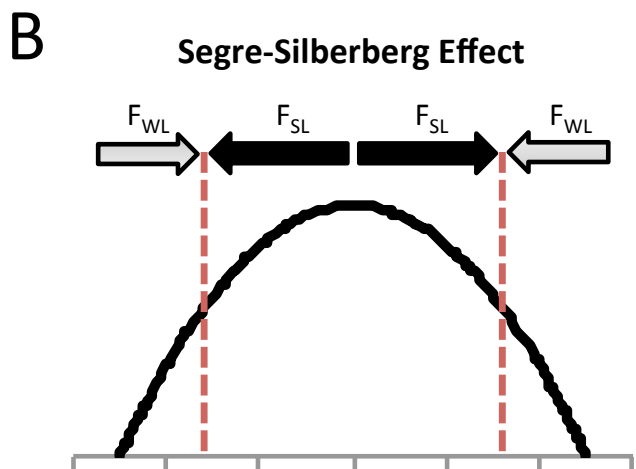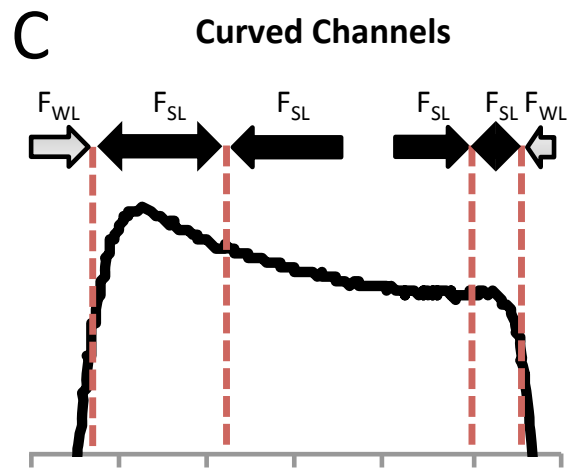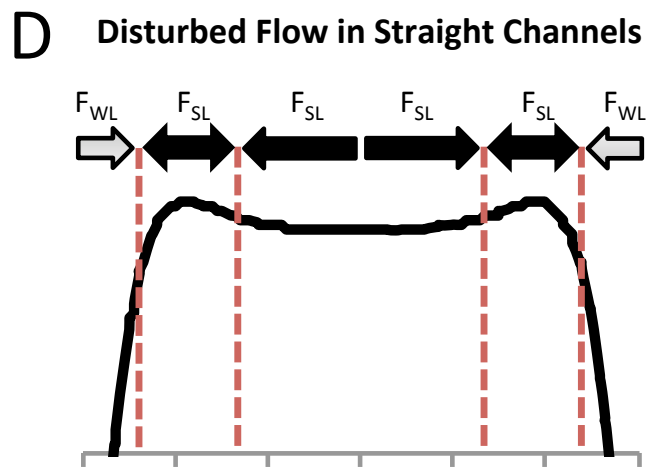

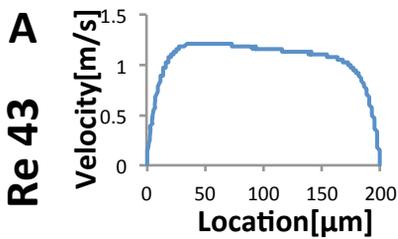
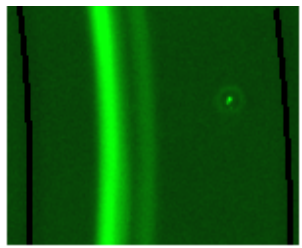
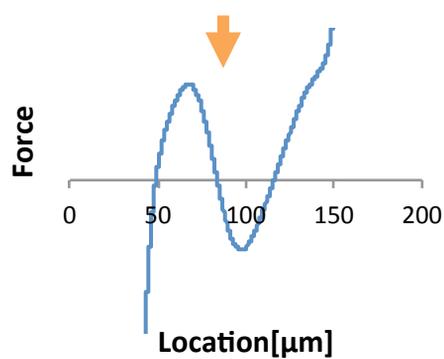
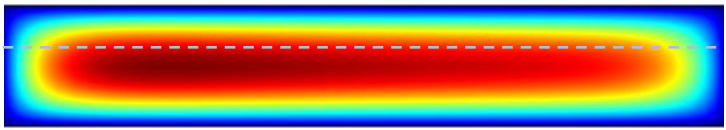
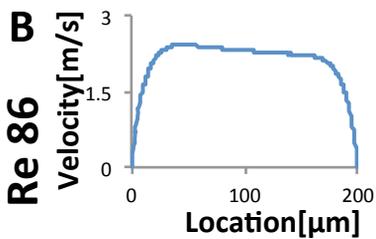
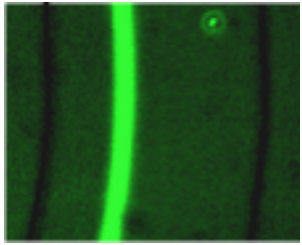
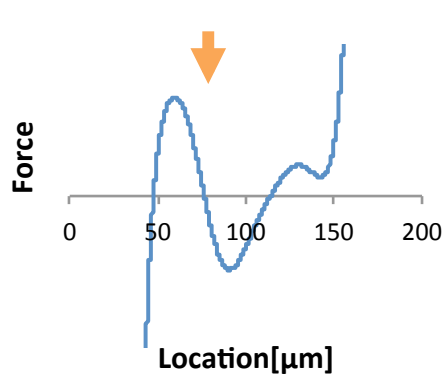
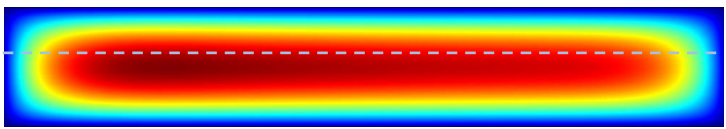
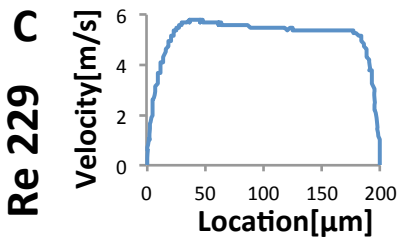
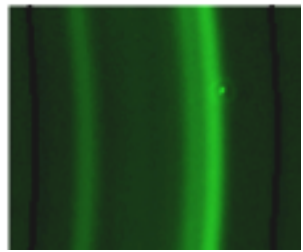
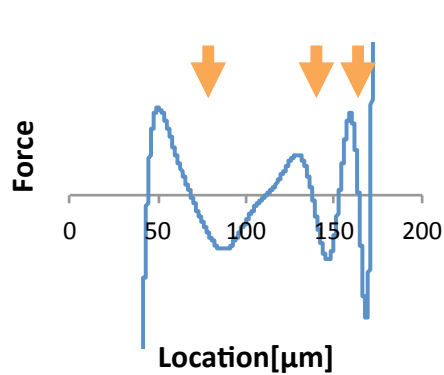
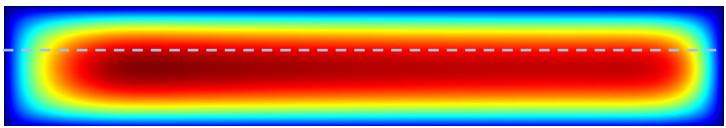
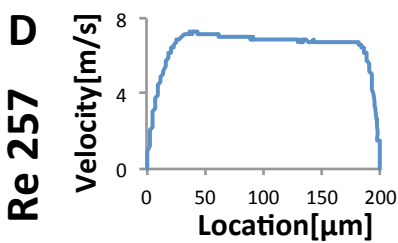
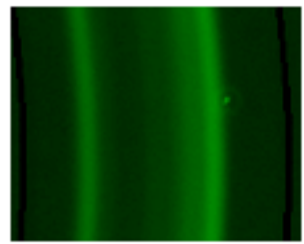
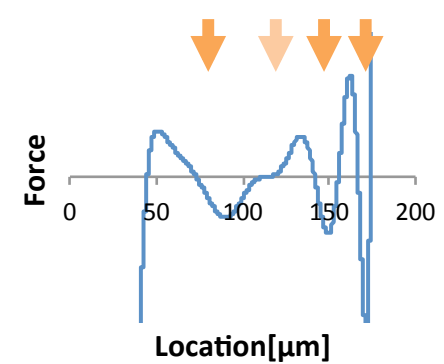
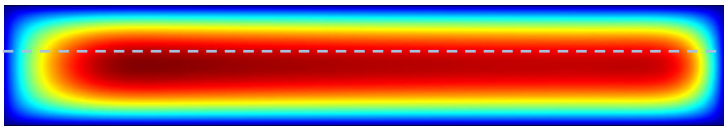

| Entry Effect Parameter Re/Re$_{crit}$ | Experimental Observation | Numerical Simulation | Analytical Force Calculations Horizontal Cross-section | | Vertical Cross-section |
|---|---|---|---|---|---|
| **A** 0.31 | 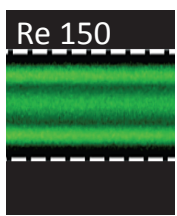 | 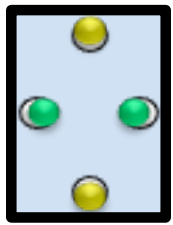 | 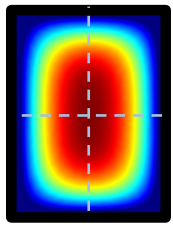 | 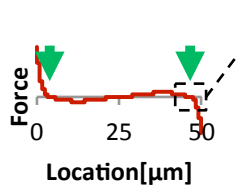 | 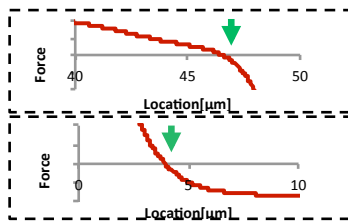 |
| **B** 0.92 | 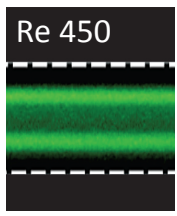 | 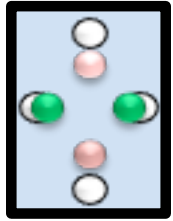 | 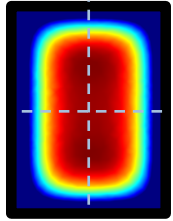 | 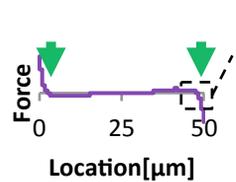 | 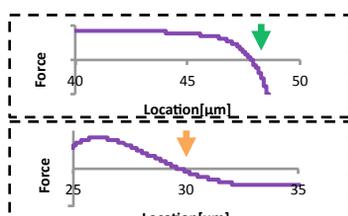 |
| **C** 1.53 | 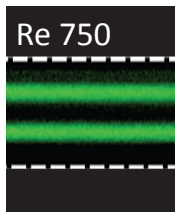 | 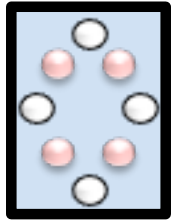 | 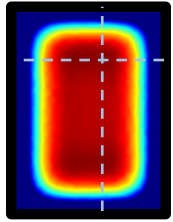 | 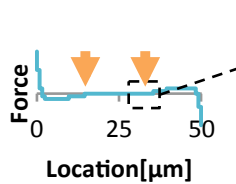 | 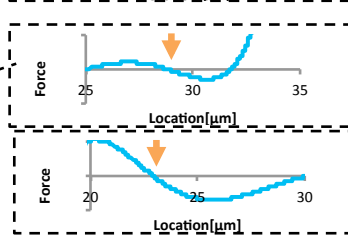 |
| **D** 2.21 | 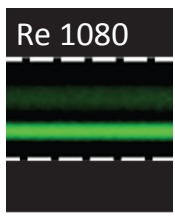 | 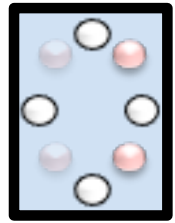 | 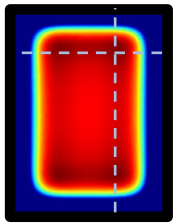 | 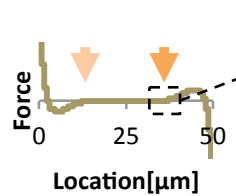 | 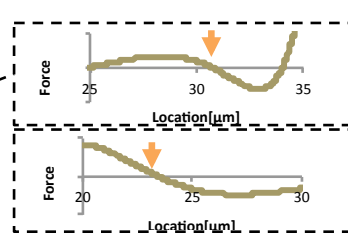 |